\documentstyle[12pt]{article}
\input epsf.tex

\begin{document}

\baselineskip=15.5pt
\pagestyle{plain}
\setcounter{page}{1}

\renewcommand{\thefootnote}{\fnsymbol{footnote}}

\begin{titlepage}

\begin{flushright}
IASSNS-HEP-98-47\\
PUPT-1794\\
Imperial/TP/97-98/47\\
hep-th/9805156
\end{flushright}
%\vspace{5 mm}
\vfil

\begin{center}
{\huge Coupling Constant Dependence in the Thermodynamics
of {\cal N} = 4 Supersymmetric Yang-Mills Theory}
\end{center}

%\vspace{2 mm}
\vfil

\begin{center}
{\large Steven S.\ Gubser}\\
\vspace{1mm}
Joseph Henry Laboratories,
Princeton University,
Princeton, New Jersey 08544, USA\\
\vspace{6mm}
{\large Igor R.\ Klebanov\footnote{On leave from 
Joseph Henry Laboratories, Princeton University.}}\\
\vspace{1mm}
Institute for Advanced Study,
Olden Lane,
Princeton, New Jersey 08540, USA\\
\vspace{3mm}
%and\\
\vspace{3mm}
{\large Arkady A. Tseytlin\footnote{Also at  Lebedev  Physics
Institute, Moscow.}}\\
\vspace{1mm}
Blackett Laboratory,
Imperial College,
London, SW7 2BZ, U.K. \\
\vspace{3mm}
\end{center}

%\vspace{1 mm}
\vfil

\begin{center}
{\large Abstract}
\end{center}

\noindent
The free energy of the maximally supersymmetric $SU(N)$
gauge theory at temperature $T$
is expected to scale, in the large $N$ limit, as
$N^2 T^4$ times a function of the 't~Hooft coupling,
$f(g_{{\rm YM}}^2 N)$. In the strong coupling limit the free energy
has been deduced from the near-extremal 3-brane geometry, and
its normalization has turned out to be $3/4$ times that found in the
weak coupling limit. In this paper we calculate the leading
correction to this result in inverse powers of the coupling,
which originates from the $R^4$ terms in the  tree level
effective action of type IIB string theory. The correction to $3/4$ is 
positive and of order $(g_{{\rm YM}}^2 N)^{-3/2}$. Thus, 
$f(g_{{\rm YM}}^2 N)$ increases as the 't~Hooft coupling is decreased, 
in accordance with the expectation that it should be approaching 1
in the weak coupling limit.
We also discuss similar corrections for other conformal theories
describing  coincident branes. In particular, we suggest
that the coupling-independence of the near extremal entropy for
D1-branes bound to D5-branes is related to the vanishing of
the Weyl tensor of $AdS_3\times S^3$.

%\vspace{1cm}
\vfil
\begin{flushleft}
May 1998
\end{flushleft}
\end{titlepage}
\newpage

\renewcommand{\thefootnote}{\arabic{footnote}}
\setcounter{footnote}{0}

%%%%%%%%%%%%%%%%%%%%%%%%%%%%%%%%%%%%%%%%%%%%%%
%% include the next line for double spacing %%
%%%%%%%%%%%%%%%%%%%%%%%%%%%%%%%%%%%%%%%%%%%%%%
%\renewcommand{\baselinestretch}{2}

\newcommand{\grad}{\nabla}
\newcommand{\tr}{\mathop{\rm tr}}
\newcommand{\half}{{1\over 2}}
\newcommand{\third}{{1\over 3}}
\newcommand{\be}{\begin{equation}}
\newcommand{\ee}{\end{equation}}
\newcommand{\bea}{\begin{eqnarray}}
\newcommand{\eea}{\end{eqnarray}}

\newcommand{\dint}[2]{\int\limits_{#1}^{#2}}
\newcommand{\D}{\displaystyle}
\newcommand{\PDT}[1]{\frac{\partial #1}{\partial t}}
\newcommand{\PD}{\partial}
\newcommand{\tw}{\tilde{w}}
\newcommand{\tg}{\tilde{g}}
\newcommand{\newcaption}[1]{\centerline{\parbox{6in}{\caption{#1}}}}
\def\href#1#2{#2}  
%%%%%%%%%%%%%%%%%%%%%%%%%%%%%%

\def \ci {\cite}
\def \foot {\footnote}
\def \bi{\bibitem}
\newcommand{\rf}[1]{(\ref{#1})}
\def \del{\partial}
\def \m {\mu}
\def \n {\nu} 
\def \g {\gamma}
\def \G {\Gamma}
\def \a {\alpha}
\def \ov {\over}
\def \la {\label}
\def \ep {\epsilon}
\def \d {\delta}
\def \k {\kappa}
\def \p {\phi}
\def \ha {\textstyle{1\ov 2}}

\def\np {  {\em Nucl. Phys.} }
\def \pl { {\em Phys. Lett.} }
\def \mpl { Mod. Phys. Lett. }
\def \prl { Phys. Rev. Lett. }
\def \pr  { {\em Phys. Rev.} }
\def \cqg { Class. Quantum Grav.}
\def \jmp { Journ. Math. Phys. }
\def\ap { Ann. Phys. }
\def \ijmp { Int. J. Mod. Phys. }

%--------+---------+---------+---------+---------+---------+---------+
%Steve's macros: these seem to work both in latex and harvmac.
%
%Macros to facilitate use of halign for complicated equations:
\def\TL{\hfil$\displaystyle{##}$}
\def\TR{$\displaystyle{{}##}$\hfil}
\def\TC{\hfil$\displaystyle{##}$\hfil}
\def\TT{\hbox{##}}
%Example: the \noalign command gives an extra bit of vertical space
%  \eqn\One{\vcenter{\openup1\jot
%    \halign{\strut\span\TL & \span\TR & \span\TT & \span\TL & \span\TR\cr
%     x^2 &> 1 & \quad when $x$ satisfies\ \ & x &> 1 \cr\noalign{\vskip1\jot}
%     y^2 &< 1 & \quad when $y$ satisfies\ \ & y &< 1 \cr
%   }}}
%seqalign takes, eg, {\span\TL & \span\TR\qquad & \span\TT} for its 
%first argument, and otherwise behaves just like eqalign.
%In order to make nice page breaks, consider using \halign to\displaywidth{...}
%instead of just \halign{}.
%See TeXBook pp. 237, and especially 193 on \eqalignno.
%Also note that \displaylines{} acts as \seqalign{\span\TC}{}.
\def\seqalign#1#2{\vcenter{\openup1\jot
  \halign{\strut #1\cr #2 \cr}}}

%Blank macros:
\def\comment#1{}
\def\fixit#1{}

%For controlling the size of fractions:
\def\tf#1#2{{\textstyle{#1 \over #2}}}
\def\df#1#2{{\displaystyle{#1 \over #2}}}

%For adding more math operators:
\def\mop#1{\mathop{\rm #1}\nolimits}

%More math operators: (add as needed)
\def\ad{\mop{ad}}
\def\coth{\mop{coth}}
\def\csch{\mop{csch}}
\def\sech{\mop{sech}}
\def\Vol{\mop{Vol}}
\def\vol{\mop{vol}}
\def\diag{\mop{diag}}
\def\tr{\mop{tr}}
\def\Disc{\mop{Disc}}
\def\sgn{\mop{sgn}}

%Group symbols:
\def\SU{{\rm SU}}
\def\USp{{\rm USp}}            

%Approximately less than operators:
\def\lsim{\mathrel{\mathstrut\smash{\ooalign{\raise2.5pt\hbox{$<$}\cr\lower2.5pt\hbox{$\sim$}}}}}
\def\gsim{\mathrel{\mathstrut\smash{\ooalign{\raise2.5pt\hbox{$>$}\cr\lower2.5pt\hbox{$\sim$}}}}}
%Used to use this:
%\def\lsim{\mathrel{\raise2pt\hbox{$\mathop<\limits_{\hbox{\raise3pt\hbox{$\sim$}}}$}}}
%\def\gsim{\mathrel{\raise2pt\hbox{$\mathop>\limits_{\hbox{\raise3pt\hbox{$\sim$}}}$}}}

%Nicest general slashing macro I can come up with:
\def\slashed#1{\ooalign{\hfil\hfil/\hfil\cr $#1$}}
%Used to use this: \def\slashed#1{\hskip2pt/\hskip-5.9pt#1} 

%To produce a box for a Dalembertian (adapted from p. 320 of TeXbook):
\def\sqr#1#2{{\vcenter{\vbox{\hrule height.#2pt
         \hbox{\vrule width.#2pt height#1pt \kern#1pt
            \vrule width.#2pt}
         \hrule height.#2pt}}}}
\def\square{\mathop{\mathchoice\sqr56\sqr56\sqr{3.75}4\sqr34\,}\nolimits}
%Extra space here looks nicer in main math text mode.

%Young Tableaux macros:
\def\idget{$\sqr55$\hskip-0.5pt}
\def\endrow{\hskip0.5pt\cr\noalign{\vskip-1.5pt}}
\def\endyoung{\hskip0.5pt\cr}
%Example: in a paragraph or in mathmode, say
%\oalign{\idget\idget\idget\idget\endrow
%        \idget\idget\idget\endyoung}
%See young.tex for more examples.

%With ssg.bst one needs this definition unless you are going to 
%set up hyperlinking.
\def\href#1#2{#2}  

%--------+---------+---------+---------+---------+---------+---------+

%These next macros allow one to type almost-harvmac in 
%a latex file.  Use \eno for forward referencing an equation.
%
\def\lbldef#1#2{\expandafter\gdef\csname #1\endcsname {#2}}
\def\eqn#1#2{\lbldef{#1}{(\ref{#1})}%
\begin{equation} #2 \label{#1} \end{equation}}
\def\eqalign#1{\vcenter{\openup1\jot
    \halign{\strut\span\TL & \span\TR\cr #1 \cr
   }}}
\def\eno#1{(\ref{#1})}

\def\rmax{{r_{\rm max}}}
\def\gone#1{}
%--------+---------+---------+---------+---------+---------+---------+

%%%%%%%%%%%%%%%%%%%%%%%%%
\section{Introduction}

A quarter of a century ago 't~Hooft showed that non-Abelian gauge
theories simplify in the limit where the number of colors, $N$,
is taken to infinity \cite{GT}. If $g_{{\rm YM}}^2 N$ is kept fixed, then
the perturbative expansion reduces to planar diagrams only.
't~Hooft speculated that these planar diagrams are to be thought of as the 
world sheets of a string. This idea has been among the dominant
themes in theoretical physics ever since.

Recently, a great deal of progress in this direction has been
taking place. In \cite{Sasha} it was suggested that the 
``confining string'' may be defined by a 
two-dimensional conformal sigma model with a 5-dimensional target space.
The 5-th dimension was argued to arise from the conformal factor of the
world sheet geometry.\footnote{A similar phenomenon is known to
occur for random surfaces embedded in 1 dimension and turns them
into a string theory with a 2-dimensional target space \cite{Polch,DJ}.
}

Another, seemingly unrelated, development is connected with the
Dirichlet brane \cite{brane} description of black 3-branes
in \cite{gkp,kleb,gukt,gkThree}.
The essential observation is that, on the one hand, the black
branes
are solitons which curve space \cite{hs,DL}
and, on the other hand, the world volume of $N$
parallel D-branes is described by supersymmetric $U(N)$ gauge theory with
16 supercharges \cite{Witten}.
A particularly interesting system is
provided by the limit of
a large number $N$ of coincident D3-branes \cite{gkp,kleb,gukt,gkThree},
whose world volume is described by ${\cal N}=4$ supersymmetric $U(N)$
gauge theory in $3+1$ dimensions. In \cite{kleb} is was shown
that their description by the 3-brane solution
of type IIB supergravity becomes reliable in the
``double-scaling limit,''
\begin{equation}
\label{dsl}
{L^4\over \alpha'^2} = 2 g^2_{{\rm YM}} N \rightarrow \infty\ ,
\qquad\qquad \omega^2 \alpha' \rightarrow 0\ ,
\end{equation}
where $L$ is the radius of the 3-brane throat, 
$ g_{{\rm YM}}^2= 2\pi g_s$,
and $\omega$ is a characteristic  excitation energy.
An equivalent statement is that $\alpha'$ is being taken to $0$.

In an important development Maldacena, with some motivation from the
results in \cite{gkp,kleb,gukt,gkThree}, from work on other
black holes in \cite{dps} and from studies of the
throat geometry in \cite{gt,ght}, made a fairly concrete ``confining
string'' proposal \cite{jthroat}. 
Taking the limit $\alpha'\rightarrow 0$
directly in the 3-brane metric, he found that the ``universal region''
describing the gauge theory is the throat, whose geometry is
the space $AdS_5\times S^5$  with  equal radii $L$ of the factors. 
The proposal \cite{jthroat} was sharpened in \cite{US,EW}
where it was shown how to calculate the correlation functions of certain 
gauge theory vertex operators from the response of the type IIB theory
on $AdS_5\times S^5$ to boundary conditions. Many other interesting
results were obtained recently, but they are outside the scope of
this paper.

An interesting ingredient the proposal \cite{jthroat}
adds to the sigma model formulation in \cite{Sasha} is the
presence of a Ramond-Ramond self-dual 5-form background field. 
Though the corresponding action for the type IIB superstring
was recently constructed in \ci{MT},   
little is known to date about properties of sigma models with Ramond-Ramond
backgrounds. In the absence of their exact solution, one can
still extract a host of useful information about strongly
coupled gauge theories by using the $\alpha'$ expansion of the
type IIB string  effective action. The study of leading $\alpha'$
corrections was recently initiated in \cite{BG}.  

The specific aspect of the gauge theory -- string theory connection that
we will pursue here is the thermodynamics of the maximally supersymmetric
$SU(N)$ gauge theory in the large $N$ limit.
In \cite{gkp} it was pointed out that the near extremal black
3-brane of Hawking temperature $T$ may be used to study
the large $N$ S{\rm YM} theory heated up to the same temperature.
The metric of the black 3-brane is
\begin{equation}
\label{metric}
   ds^2 = 
H^{-1/2}(r)
    \left[ - f(r) dt^2 + d\vec{x}^2 \right] +
H^{1/2}(r)
    \left[f^{-1} (r) dr^2 + r^2 d\Omega_5^2 \right] \ ,
\end{equation}
where 
$$   H(r)  = 1 + {L^4 \over r^4} \ , \qquad \  \  f(r) = 1- {r_0^4\over r^4}
\ .
$$
The horizon is located at $r=r_0$ and the extremality
is achieved in the limit $r_0 \rightarrow 0$ where the Hawking
temperature vanishes. In order to study the conformal limit of
the world volume theory of coincident 3-branes
we need to keep $TL \ll 1$. This leads to the requirement that
the metric is near-extremal, i.e.   $r_0\ll L$. Here the
Hawking temperature
\begin{equation}
\label{Hawking}
T = {r_0\over \pi L^2}\  \ll \  { 1 \ov L}
\ .
\end{equation}

In \cite{gkp} the entropy of the $SU(N)$ 
S{\rm YM} theory was identified with the 
Bekenstein-Hawking entropy of the geometry (\ref{metric}).
The result turned out to have a remarkable form,
\begin{equation}
\label{bhe}
S_{BH}= {\pi^2\over 2} N^2 V_3 T^3 
\ .
\end{equation}
The $T^3$ scaling is in agreement with the conformal invariance of
the theory. Indeed,  the extensivity of the entropy, and the
absence of another scale in the theory, requires the $V_3 T^3 $
dependence. Furthermore, the factor $N^2$ indicates that one is 
dealing with a theory of $N^2$ unconfined massless degrees of freedom.
As in \cite{gkp}, it is instructive to compare (\ref{bhe})
with the entropy of the weak coupling limit of the S{\rm YM}
theory, where it reduces to that of $8 N^2$ free massless bosons and fermions.
Now the answer turns out to be
\be S_0= {2 \pi^2\over 3} N^2V_3 T^3 
\ .
\ee
Thus,\  $ S_{BH}= {3\over 4} S_0$.

 Is there a definite contradiction here?
The answer is no, because $S_{BH}$ is relevant to the S{\rm YM}
theory in the limit of $N\to \infty$ {\it and large 't~Hooft coupling}
$g_{{\rm YM}}^2 N$.  The relations $L^4/\ell_{\rm Planck}^4 \sim N$ and
$L^4/\ell_{\rm string}^4 \sim g_{{\rm YM}}^2 N$ reveal that only in this
limit are both the $\alpha'$ and the loop corrections to the solution
(\ref{metric}) negligible.
In this paper
our focus will be on the $\alpha'$ corrections.  We suppress the loop
corrections, which amounts in the gauge theory to taking the
$N\to\infty$ limit first with fixed 't~Hooft coupling. 

On general grounds, one expects the following expression for the
entropy of the large $N$ gauge theory, 
\begin{equation}
S (g_{{\rm YM}}^2 N) =\  f(g_{{\rm YM}}^2 N)\ {2 \pi^2\over 3} N^2V_3 T^3 
\ ,
\end{equation}
where $f(g_{{\rm YM}}^2 N)$ is (hopefully) a
smooth function. Its weak coupling limit
is $f(0)= 1$. The supergravity approach \cite{gkp}
predicts that, in the limit
$g_{{\rm YM}}^2 N \rightarrow \infty$, $f$ should  approach $3\ov 4$. 

In order to study precisely how this limiting value is
approached, it is necessary to take into
account the leading higher derivative correction in the
type IIB tree level effective action. As discussed in
\cite{US,EW} the corrections to the  effective 
action in the $AdS_5\times S^5$ background
are measured by the dimensionless parameter
\be {\alpha'\over L^2} = (2 g_{{\rm YM}}^2 N)^{-1/2}
\ .
\ee
Thus, the $\alpha'$ expansion in the type IIB theory
translates into the strong 't Hooft coupling 
expansion in the S{\rm YM} theory.\foot{The string loop effects give
rise to corrections of order $g_s^2 \alpha'^4/L^8 \sim 1/N^2$.
Work on such corrections to the entropy was 
initiated in \cite{HO,BR}.}
This will help us identify the leading correction.
Our main result is that the strong coupling expansion for $f$ is
\be
 f(g_{{\rm YM}}^2 N) = {3\over 4} + 
{45\over 32} \zeta(3) (2 g_{{\rm YM}}^2 N)^{-3/2}  + \ldots
\ .
\ee
In section~\ref{FreeEnergy} we derive this result by first order
perturbation theory on the supergravity action.  A priori one might
expect this direct treatment to be inadequate because it ignores
perturbations in the geometry which change, for instance, the relation
between the temperature and the non-extremality parameter $r_0$.  We
show in section~\ref{PerturbedSolution}  that taking the
perturbations to the geometry into account does not, however, 
 change the final answer for the free energy to first order.

Section~\ref{OtherCFTs} summarizes our
results for other CFT's  describing coincident branes.
 In
\ci{ENT}, the entropy of the non-dilatonic near-extremal p-branes
(i.e. D3, M2, M5 branes) was computed from the leading-order
supergravity solution.  The free energy has the form expected of a
CFT in $p$ dimensions:
 \be
F = - k_p  N^{p+1 \ov 2} V_p T^{p+1}   \ ,
\la{eee}
\ee
where $N$ is the quantized p-brane charge or the number of 
coincident branes, and $k_p$ is a constant of order unity. 
The conformal invariance of the corresponding world-volume
theories fixes the temperature dependence, so that the higher-order
corrections to the supergravity action are expected to change only the
$N$-dependent coefficient by subleading terms. 
  In particular, 
we derive the  first $1/N$ corrections to the free energies in the case of 
M5-branes and M2-branes, generalizing the  
leading-order results of \ci{ENT}. 

For the near-horizon
geometry which describes D1-branes bound to D5-branes at
strong coupling we find that the leading $\alpha'$ correction
to the free energy vanishes. This is due to the fact
that the leading $\alpha'$ correction may be expressed in terms
of the Weyl tensor \cite{BG}, while the geometry is
locally $AdS_3\times S^3$ which is conformally flat
so that the Weyl tensor vanishes.
Based on this, we further argue that there are no corrections to
all orders in $\alpha'$.
This seems to explain why the near-extremal entropy
is independent of the coupling in this case \cite{cm,HSS}.

%%%%%%%%%%%%%%%%%%%%%%%%%%%%%%%%%%%%%%%%%%%%%%%%%%%%%%
\section{The free energy from the gravitational action}
\label{FreeEnergy}
%%%%%%%%%%%%%%%%%%%%%%%%%%%%%%%%%%%%%%%%%%%%%%%%%%%%%%%%%%%%%

Since the horizon of a near-extremal 3-brane is located far down its
throat, the same answer for the Bekenstein-Hawking entropy
is obtained if we replace the 3-brane metric by the throat approximation,
$r \ll L$. The resulting metric \ci{jthroat,HR}
\begin{equation}
\label{throatmetric}
   ds^2 = {r^2 \over L^2} 
    \left[ -  (1- {r_0^4\over r^4})
dt^2 + d\vec{x}^2 \right] +
    {L^2 \over r^2} 
    ( 1- {r_0^4\over r^4})^{-1}  dr^2 + L^2 d\Omega_5^2 \ ,
\end{equation}
is a product of $S^5$  with  a  certain
limit of the Schwarzschild black hole in $AdS_5$
 \cite{newWit}. The Euclidean Schwarzschild black hole is asymptotic to
$S^1\times S^3$, and the required limit is achieved as the volume of
$S^3$ is taken to infinity. Thus, the Euclidean continuation of
the metric (\ref{throatmetric}) is asymptotic to $S^1\times R^3$, 
and the circumference
of $S^1$ is $\beta=1/T$. 
Indeed, if we  set $r= r_0(1  + L^{-2} \rho^2)$ near $r_0$, 
then the relevant 2d  part of the Euclidean  metric becomes:
\begin{equation}
ds^2 = d\rho^2 + {4r_0^2\ov L^4} \rho^2 d\tau^2\ , \ \  \qquad \tau=it \ ,
\end{equation}
so that the   period of the Euclidean time
 required by the regularity of the metric is
$\beta = \pi L^2/r_0$.

Following \ci{HP}, one can
identify the free energy $F$ of the theory with the Euclidean 
 gravitational
action times the temperature, i.e. 
\be  I= \beta F   \ . \ee
 The 5-dimensional gravitational action 
obtained  from the $D=10$ type IIB supergravity action
by compactifying on $S^5$ 
is 
\be
 I_5 = -{1\over 16\pi G_5} \int d^5 x \sqrt{g_5} \left(R_5 + {12\over L^2} 
\right) \ .
\la{lead}
\ee
Calculation of the contributions from the 2-derivative terms is divergent
at large distances and requires a subtraction. The end result
is that, as expected, the entropy is the horizon area divided by
$4 G_5$  \ci{newWit}. The leading term in the free energy is
\be F_0= - {\pi^2\over 8} N^2 V_3T^4 \ . 
\ee

We will focus on the contribution to $F$ coming 
 from the $\a'^3 R^4$ string  correction \ci{GZ,GW}
to the supergravity  action.
In the Einstein frame, and using the convention of including
$(F_5)^2$ in the action and imposing self-duality after the equations
of motion are derived, the tree level type IIB  string effective action
 has the following structure: 
\be
I=
 -{1\over 16\pi G_{10}} \int d^{10} x \sqrt g
\ \bigg[ R - \half (\del \phi)^2 - {1 \ov 4 \cdot 5!}   (F_5)^2  +...+ 
\  \g \ e^{- {3\ov 2} \phi}  W + ...\bigg]   \  ,
\la{aaa}
\ee
$$ \ \ \ \ \ \   
  \g= { 1\ov 8} \zeta(3)(\alpha')^3 \ , 
$$  
$$
W = R^{hmnk} R_{pmnq} R_{h}^{\ rsp} R^{q}_{\ rsk} 
          \ + \ \half  R^{hkmn} R_{pqmn} R_h^{\ rsp} R^{q}_{\ rsk}  
$$
\be
+ \ {\rm terms\  depending\  on\  the\ Ricci\  tensor} \  . 
\la{rrer}
\ee
Dots stand for other terms depending on  antisymmetric tensor
field strengths and 
derivatives  of dilaton.\foot{As follows
 from  \ci{GZ}, there exists a  scheme
in which the  string-frame  metric and dilaton 
dependent terms in the  $O(\a'^3)$ action 
are given by 
$\int d^{10} x \sqrt {G} \ e^{-2\p} 
\ \big[ R  + 4(\del \phi)^2  +  
  \g W(R)\big], $ where 
$W(R)$ is the Riemann tensor part of  $W$.}
The field redefinition ambiguity \ci{GW,TF}
allows one to change the coefficients of terms 
involving the Ricci tensor (in essence, ignoring other fields,  
one may use $R_{mn}=0$ to simplify the structure of $W$  as
the graviton legs in the 4-point  string amplitude are on-shell).
Thus there exists a scheme where
$W$ depends only on the Weyl tensor
\be
W =  C^{hmnk} C_{pmnq} C_{h}^{\ rsp} C^{q}_{\ rsk} 
 + \half  C^{hkmn} C_{pqmn} C_h^{\ rsp} C^{q}_{\ rsk}  
 \  . 
\la{rrrr}
\ee
In general, there are other terms of the same order
which accompany $C^4$ by supersymmetry. 
The Riemann tensor dependent part of  $W$ 
(the first line in \rf{rrer}) may be written as
\be W= {1\ov 3 \cdot 2^8}  J_0 \ , \ \ \ \ \ \  \ \    
J_0= t_8\cdot t_8 RRRR - {1\ov 4} \ep_{8}\cdot \ep_{8} RRRR
\ ,
\la{kkk}
\ee
which is the bosonic part of a superinvariant 
 \ci{NT,SUL,TRR}.\foot{Note that while the
 on-shell N=1, D=10  superinvariant  \ci{NT,BG} 
$\int  d^{10} x d^{16} \theta \  \Phi^4  \to 
\int d^{10} x d^{16} \theta \ (\bar \theta  \gamma^{mnk} 
\theta\bar \theta  \gamma^{pq}_{\ \ \  k}
\theta  R_{mnpq})^4$
depends  only on the Weyl tensor   (because of the identity  $
 \gamma^{mnk} \theta\bar  \theta \gamma_{mnl}\theta 
\equiv 0$)  and is proportional to $W$ in \rf{rrrr}, 
its off-shell extension   \rf{kkk} contains the  Riemann tensor 
\ci{SUL}, i.e. it should involve  
also the Ricci   tensor or derivatives
of the dilaton and other fields.}
Here the $t_8\cdot t_8 R^4$ term has the structure 
24 STr$[R^4-{1\ov 4} (R^2)^2]$
%(i.e. does not include $\ep_8$ term)  
while the $\ep_{8}\cdot \ep_{8} RRRR$ term is defined in 
$D$-dimensional Euclidean space as 
$$ {1 \ov (D-8)!} \ep_{_D}\cdot \ep_{_D} RRRR=
8! 
\delta^{n_1}_{[m_1}
\ldots \delta^{n_8}_{m_8]} R^{m_1 m_2}_{n_1 n_2} \ldots R^{m_7 m_8}_
{n_7 n_8}
\ .
$$

The  choice \rf{rrrr}  
of $W$ is special in that the leading-order 
$AdS_5 \times S^5$+$F_5$-strength  solution  \ci{SH},
which has a conformally flat metric, 
is then not modified by the $R^4$ correction \ci{BG}. 
Strictly speaking, the above discussion does not rule out 
that the additional  $F_5$-dependent terms (ignored in \ci{BG})
may  lead  to a modification of the solution, but  more general
arguments  based on maximal supersymmetry of this  
background \ci{MT,KR} suggest that this does not happen.\foot{As in 
the case of the group space compactification  (the WZW model), 
the  special scheme choice is crucial also for 
 having the parameters of $AdS_n \times S^m$  backgrounds being 
unchanged by higher-order corrections (for example, the
field redefinitions like
$g_{mn} \to g_{mn} + a R_{mn} + ... $
rescale the factors in the metric).}

The non-extremal  background  with the metric  \rf{throatmetric}
(and constant dilaton $\p$  and self-dual $F_5$ field being the same as 
in the  extremal case $r_0=0$) 
will, however,  suffer a  modification,   
just like the Schwarzschild  solution is modified 
by $R^2$ corrections present  in the bosonic or heterotic
string theory \ci{CMP}.
We will discuss the detailed form of this modification in the next
section.

In this section we will use a simpler approach, which is essentially
the traditional first order perturbation theory for the
free energy. To evaluate the 
leading correction to the free energy, we will substitute 
the unperturbed metric
\rf{throatmetric} into the action term $W$. 
A more detailed calculation
in the next section provides a check on this simple procedure.
Possible  extra $F_5$-dependent 
terms  should  not affect our  computation  
since the 5-form field  for the non-extremal  background 
\rf{throatmetric} is the same as in the case of $r_0=0$.
The shift of the dilaton  from its constant  value
(which we  choose to be $\p_0=0$)
changes the value of the action only at the 
next order of perturbation theory in $\gamma$.

The  `$AdS_5$'  part of the (Euclidean)  metric \rf{throatmetric}
has the following Ricci and Weyl tensors
($m,n=\tau,r,1,2,3; \ a,b=\tau,r; \ i,j,k=1,2,3$)
\be
R_{mn} = -{ 4\ov L^2} g_{mn} \ , \ \ \   \ \ \  \ \ R= -{ 20\ov  L^{2}}  \ , 
\ee
$$
C^{ab}_{\ \ cd} =  3 X  \epsilon^{ab} \epsilon_{cd}\ , \ \ \ 
C^{ai}_{\ \ bj} =  - X  \delta^a_b \delta^i_j \ , \ \ \ 
C^{ij}_{\ \ kl} =    X ( \d^i_k \d^j_l - \d^j_k \d^i_l) \ , 
\ \ \ \
X\equiv { 1\ov L^2} {  r_0^4\ov  r^4} \ , 
$$
so that  $W$ in  \rf{rrrr} is given by\foot{The same result is found for
 the full 10-dimensional  metric \rf{throatmetric}. This is related
to the fact that the the value of the Ricci scalar of this metric is zero
in both the extremal and the non-extremal cases.}
\be
W=\   { 180  \ov L^8}\  {r_0^{16} \ov  r^{16} } \ . 
\la{wel}
\ee
The simplicity of this 
result  is obviously a consequence of the conformal  flatness of the $AdS_5$ 
metric which is the $r_0\to 0$ 
 limit of \rf{throatmetric}.  
As a result, 
the correction to the action 
\begin{equation}
\delta I= {T}^{-1}  \delta  F 
 =-{1\over 16 \pi G_5} \int d^5 x \sqrt {g_5 }\  \g \  W\ 
\end{equation}
 is 
a perfectly convergent integral at large $r$.
Since Vol$(S^5) = \pi^3 L^5$, we have
\begin{equation}
{1\over 16 \pi G_5} = {\pi^3 L^5\over 16 \pi G_{10}}=
{\pi^3 L^5 \over 2 \kappa^2 }\ .
\end{equation}
Thus, using $\sqrt {g_5} = ({r\ov L})^3$, we arrive
at the following form of the correction to the free energy, 
\begin{equation}
\delta F = - {\pi^3 L^5 \over 2 \kappa^2 } V_3 {\alpha'^3\over L^{11}}
{45\over 2} \zeta(3) \int_{r_0}^\infty dr\ r^3 \ {r_0^{16}\over r^{16}}
\ .
\end{equation}
For the unperturbed solution, the temperature is
$T = {r_0\over \pi L^2}$.
Combining this with the relation based on the charge
quantization rule for 3-branes \cite{gkp,US},
\begin{equation}
\qquad  L^4 = {N\kappa\over 2 \pi^{5/2}} = 2 g_{{\rm YM}}^2 N \alpha'^2
\ ,
\end{equation}
we find
\begin{equation}
\delta F= \ - {\pi^2\over 8} N^2 V_3 T^4 \ 
{15\over 8} \zeta(3) (2 g_{{\rm YM}}^2 N)^{-3/2} 
\ .
\ee
Thus,
\be F = F_0+ \delta F = - {\pi^2\over 8} N^2V_3 T^4 
\left [ 1+ {15\over 8} \zeta(3) (2 g_{{\rm YM}}^2 N)^{-3/2} \right
] \ , \la{coo}
\ee
so that the leading correction is {\it positive}.

Thus, if  we write
\be
 F=   -\  f(g_{\rm YM}^2 N)\ {\pi^2\over 6}
  N^2 V_3T^4  \ ,
\ee
so that $f$ approaches 1 for small $g_{{\rm YM}}^2 N$, then for large
$g_{{\rm YM}}^2 N$ we have
\be f(g_{{\rm YM}}^2 N) = {3\over 4} + {45\over 32} \zeta(3) \ 
(2 g_{{\rm YM}}^2 N)^{-3/2}  + \ldots
\ .
\la{voot}
\ee

%%%%%%%%%%%%%%%%%%%%%%%%%%%%%%%%%%%%%%%%%%%%%
\section{The perturbed solution}
\label{PerturbedSolution}
%%%%%%%%%%%%%%%%%%%%%%%%%%%%%%%%%%%%%%%%%%%%%%%%%%%%%%%%

In order to find the perturbed solution, we need to generalize
the calculation of the action to arbitrary static
 metrics with same symmetry  properties. 
This turns out to be
surprisingly simple.  If we  consider  the  $(p+2)$-dimensional 
metric 
  \eqn{GenMet}{
   ds^2 = H^2 \left( K^2 d\tau^2 + P^2 dr^2 + \sum^p_{i=1} d{x}^2_i \right)
  }
 where $H$, $K$, and $P$ are functions of $r$ only, then
the invariant in \rf{rrrr} is given by\foot{Let us note that 
for this class of metrics $W\sim (C_{mnkl} C^{mnkl})^2$, and
$$C_{mnkl} C^{mnkl}=
q_{p+2} { 1 \over K^2 H^4 P^2}
  \bigg[ \bigg( 
 {K' \over P} \bigg)'\bigg]^2\ , \ \ \ \ \ 
q_4 = { 2}, \   \ q_5 = {4\ov 3}  ,     \ \
\   q_7 = { 8 \ov 3}\ . $$}
 \eqn{GenW}{
  W = d_{p+2} { 1 \over K^4 H^8 P^4}
  \bigg[ \bigg( 
 {K' \over P} \bigg)'\bigg]^4 \ ,  }
where for the most interesting  cases of $p+2=4,5,7$
$$
d_4 = { 1 \ov 18} \ , \ \ \ \  \ \  d_5 = {5\ov 36} \ , \ \ \ \    
\ \  d_7= { 292 \ov 1125} \ .
  $$
 Primes will always denote derivatives with respect to $r$.

Because of the  symmetry of the metric, we may reduce the 
action \rf{aaa} to one dimension:
  \eqn{SOne}
{
   I = -{N^2 \over 8 \pi^2} V_3 \beta \int_{r_0}^\infty dr \, 
      \sqrt{g_5}\  \left(R_5 + 12  - 
       {1 \over 2} (\partial_r \phi)^2 + 
      \g e^{-{3\over 2} \phi} W \right)
 \ ,  }
 where $\g = {1 \over 8} \zeta(3) \alpha'^3$.  In this section we shall 
set $L=1$ (the dependence on $L$ of  $\gamma$-dependent corrections
can be restored  by $\gamma \to \gamma L^{-6}$).  

 There is a slight subtlety in the derivation of \SOne: the 5-form
field strength is required to be self-dual in the full,
$\alpha'$-corrected, ten-dimensional metric; and as in the unperturbed
solution, there are $N$ units of 5-form flux through the $S^5$.
  The condition of self-duality is easily made explicit
when the ten-dimensional 
metric has no components mixing the $S^5$ and the $AdS_5$ directions, 
which is
the case even when the $\alpha'$ corrections are taken into account. 
Then for each of the 5-dimensional factors we have
$F_{abcde}=\sqrt {h_5} \epsilon_{abcde}$, where
$h_{5ab}$ is the 5-dimensional part of the metric, 
and there  are no `mixed' components 
of the 5-form field strength. This choice
solves the 5-form equations of motion.
Moreover, the corrections  to these equations 
coming from possible derivative  $DF_5$ 
terms which accompany the $R^4$ terms  also vanish.
Using the ten-dimensional Einstein equation, one can verify that, 
with this choice of the 5-form background, the
cosmological constant in \SOne\ receives no explicit $\alpha'$
corrections.\foot{ 
We are grateful to E. Kiritsis for raising the question about 
our  solution that prompted us to add 
the above explanation.}

 To leading order in $\g$  the  dilaton perturbation
$\p_1=\p-\p_0$ enters the action as 
\be \la{dil}
   I(\p_1)   = -{N^2 \over 8 \pi^2} V_3 \beta \int_{r_0}^\infty dr \, 
      \sqrt{g_5}\  \left( - \ha \p'^2_1 - 
{\textstyle {3\ov 2}} \g  \p_1 W + ... \right)
 \ .  \ee
 To this order the dilaton perturbation does not mix with the metric
perturbation, so the dilaton can be ignored altogether in the
computation of the correction to the action.

  To express the
action \rf{SOne}
most simply, it helps to set
  \eqn{MetComps}{
   H = r\  ,  \qquad K = e^{a+4b}\ ,  \qquad P = e^b \ .
  }
 Then we can write
  \eqn{EllWDef}{\eqalign{
   \ell &= \sqrt{g_5}\ (R_5 + 12) =
    -2 r e^{a+3b} (2 + r a') + 12 r^5 e^{a+5b} - 
    2 {d \over dr} 
\left[ (a'+4b') r^3 e^{a+3b} \right] \,, \cr
   w &= \sqrt{g_5}\ W = {5 \over 36} {e^{a-3b} \over r^3}
    \left( a'^2 + 7 a'b' + 12 b'^2 + a'' + 4 b'' \right)^4 \,, \cr
   I &= -{N^2 \over 8 \pi^2} V_3 \beta \int_{r_0}^\infty dr 
    \left[ \ell(a,a',b) + \ \g\ w(a,a',a'',b,b',b'') \right] \ .
  }}
 The Euler-Lagrange equations which follow from this action,
  \eqn{ELeqs}{\eqalign{
   {\partial\ell \over \partial a} - 
    {d \over dr} {\partial\ell \over \partial a'} &= 
    -\g \left( {\partial w \over \partial a} - 
     {d \over dr} {\partial w \over \partial a'} +
     {d^2 \over dr^2} {\partial w \over \partial a''} \right)\,,  \cr
   {\partial\ell \over \partial b} &= 
    -\g \left( {\partial w \over \partial b} - 
     {d \over dr} {\partial w \over \partial b'} +
     {d^2 \over dr^2} {\partial w \over \partial b''} \right) 
  }}
 can be manipulated into quite tractable equations, namely
  \eqn{SepEqs}{\eqalign{
   b' + (2 r^3 + \g w_a) e^{2b} &= 0  \,,\cr
   a' + {2 \over r} - (10 r^3 + \g w_b) e^{2b} &= 0\,, 
  }}
 where 
  \eqn{wawb}{\eqalign{
   w_a &= {1 \over 6 r^2 e^{a+5b}} 
    \left( {\partial w \over \partial a} - 
     {d \over dr} {\partial w \over \partial a'} +
     {d^2 \over dr^2} {\partial w \over \partial a''} \right)  \cr
    &= 90 {r_0^{12} \over r^{13}} (-16 r^4 + 19 r_0^4) + O(\g) \,, \cr
   w_b &= {1 \over 6 r^2 e^{a+5b}} 
    \left( {\partial w \over \partial b} - 
     {d \over dr} {\partial w \over \partial b'} +
     {d^2 \over dr^2} {\partial w \over \partial b''} \right)  \cr
    &= 90 {r_0^{12} \over r^{13}} (-64 r^4 + 79 r_0^4) + O(\g) \ .
  }}
 The first equation in \SepEqs\ is separable when $w_a$ is regarded as
a function of $r$ only.  We may take $e^{-2b} \to 0$ at the horizon as
a boundary condition on $b$ (then the position of the horizon $r=r_0$ is not shifted).  Thus
  \eqn{GotExpB}{
   e^{-2b} = \int_{r_0}^r ds \, \left[ 4s^3 + 2 \g w_a(s) \right] \ ,
  }
 and the second equation in \SepEqs\ can be integrated directly.  The
result is
  \eqn{GotAB}{\eqalign{
   b &= -\tf{1}{2} \log (r^4 - r_0^4) +  \tf{15}{ 2}\g
    \left( 5 {r_0^4 \over r^4} + 5 {r_0^8 \over r^8} - 
     19 {r_0^{12} \over r^{12}} \right) + O(\g^2) \,, \cr
   a &= -2 \log r + \tf{5}{2} \log (r^4 - r_0^4) - \tf{15}{ 2}\g 
    \left( 25 {r_0^4 \over r^4} + 25 {r_0^8 \over r^8} - 
     79 {r_0^{12} \over r^{12}} \right) + O(\g^2) \ .
  }}
 There is an arbitrary choice of additive constant in the $O(\g)$
term in $a$, corresponding to rescalings of the time variable.  The
choice made in \GotAB\  
leads to the same normalization of the time
variable as in the extremal $AdS_5$ metric.  Different normalizations of
the time variable, $t \to \lambda t$, change the final relation
\eno{GotFT} by sending $F \to \lambda F$, $T \to \lambda T$.  Clearly, 
our choice of normalization is preferred once one chooses the
particular timelike Killing vector field $\partial/\partial t$ in the
extremal solution.

For completeness let us note that the
equation for the dilaton perturbation,
\be
{ 1 \ov r^{3}}  \left[r (r^4 - r^4_0) \p_1'\right]' = {{3\ov 2}} \g W 
,\ee
has the following solution regular at the horizon, 
\be
\p_1= - { {45  
 \ov 8 }} \g  \left({   r^4_0 \ov  r^4  } + {   r^8_0 \ov 2 r^8  } +
 {  r^{12}_0 \ov 3  r^{12}}\right)
\ . 
\ee
Thus, the leading $\alpha'$ correction
makes the dilaton and  the effective string coupling $e^\phi$
decrease  towards the horizon. 

The temperature is defined through the absence of a conical
singularity in the periodic Euclidean metric: the surface gravity at
the horizon is
  \eqn{FindKappa}{
  {\hat  \kappa}  = 2 \pi T = \sqrt{g^{rr}} {d \over dr} \sqrt{g_{00}} 
    \Bigg|_{\rm h} = 
    e^{a+3b} \left( a' + 4b' + {1 \over r} \right) \Bigg|_{r=r_0} = 
    2 r_0 (1 + 15 \g) \ .
  }
 We may now compute the $O(\g)$ correction to the free energy as
a function of the temperature by calculating the gravitational action of
the perturbed geometry and then using \FindKappa.
The integral defining the action, $I$, must be
regulated by subtracting off its zero temperature limit, $I_0$.
Following \cite{newWit} we accomplish this by first cutting off the
integrals at a large radius $\rmax$.  Thus, to leading nontrivial
order in $\g$,
  \eqn{SandSzero}{\eqalign{
   I &= -{N^2 \over 8 \pi^2} \beta V_3 \int_{r_0}^\rmax dr \,
    \sqrt{g_5}\  (R_5 + 12 + \g W)  \cr 
    &= {N^2 \over 4 \pi^2} \beta V_3 (r_{\rm max}^4 - r_0^4) 
     \bigg( 1 - 75 \g \bigg[ {r_0^4 \over r_{\rm max}^4} + 
      O\bigg( {r_0^8 \over r_{\rm max}^8} \bigg) \bigg] \bigg) \,, \cr
   I_0 &= -{N^2 \over 8 \pi^2} \beta' V_3 \int_0^\rmax dr \,
    \sqrt{g_5}\ (R_5 + 12)  \cr
    &= {N^2 \over 4 \pi^2} \beta' V_3 r_{\rm max}^4 \ .
  }}
 In the expression for $I_0$ we have noted that the Weyl tensor term
vanishes, and we have defined the periodicity 
  \eqn{WhichBetaPrime}{
   \beta' = \beta e^{a+4b}\Big|_\rmax = 
    \beta \left[ 1 - \tf{1}{2} (1 + 75 \g  ) {r_0^4 \over r_{\rm max}^4} 
     + O\left( {r_0^8 \over r_{\rm max}^8} \right) \right] 
  }
 for the Euclidean time so that the proper length of the circle which
it parametrizes is the same at the radius $\rmax$ as for the
near-extremal metric.  After taking the difference of $I$ and $I_0$ we
may send $\rmax \to \infty$ thus finding the following expression
for the free energy, 
  \eqn{GotFT}{
   F = {I - I_0 \over \beta} = -{N^2 \over 8 \pi^2} V_3 r_0^4 
     (1 + 75 \g ) = -{\pi^2 \over 8} N^2 V_3 T^4 (1 + 15 \g ) \ .
  }
 Surprisingly, this is equal 
(after $\gamma \to \gamma L^{-6}$) to the result  \rf{coo} obtained in
the previous section by ignoring all
perturbations to the metric and computing only the change in the
action from the Weyl tensor term.  That means that the features of the
geometry relevant to its thermodynamics are miraculously unchanged at
order $\alpha'^3$. 
Notice that
the ratios $T/r_0$, $\beta'/\beta$, and the action $I$ have all
changed from the values used in section~\ref{FreeEnergy}.  But the
basic thermodynamic relation \eno{coo} remained the same.

Using standard thermodynamics, it is easy to derive the
 corresponding expression for the entropy. It 
 has the same $\g$ dependent factor as the free energy,
\be
S= {\pi^2 \over 2} N^2 V_3 T^3 (1 + 15 \g)
\ .
\ee
Let us observe that the $O(\g)$ correction to the
entropy cannot be obtained by evaluating the horizon area of the
perturbed solution divided by $4 G_5$. 
The latter gives instead
\be
{A_h\over 4 G_5} = {\pi^2 \over 2} N^2 V_3 T^3 (1 - 45 \g)
\ .
\ee
The fact that the entropy is not directly related to the 
horizon area in higher-derivative gravity was
already noted in \cite{CMP,Myers,MinStew,Koga}.

So far we have discussed the corrections to the entropy in inverse
powers of the 't~Hooft coupling, which originate from the 
$\alpha'^n$ corrections in the tree level
string effective  action. Another interesting direction is to consider
the $1/N^2$ effects 
which originate from the string loop corrections. In fact, the leading term 
induced at one loop is of the same $R^4$  form  \rf{rrer}  that determines the leading
planar correction.
Following  \ci{BG}
 we can actually replace the tree-level $\zeta(3)$-coefficient of the $R^4$ term 
 by the function \ci{GG} containing tree-level, one-loop
and non-perturbative D-instanton  corrections 
$$
2\zeta(3) (2 g^2_{\rm YM} N)^{-3/2} \ \ \to \ \ 
2\zeta(3) (2 g^2_{\rm YM} N)^{-3/2} + { 1 \ov 24N^2} 
(2 g^2_{\rm YM} N)^{1/2}
$$ \be \ \ \ \ \ \  
+\  {1 \ov N^{3/2} }\ h(e^{-4\pi^2/g^2_{\rm YM}})\ (1 + o(g^2_{\rm YM})) \ , 
\ee
where 
$h$ represents  infinite series of instanton corrections
(our definition of $g^2_{\rm YM}= 2 \pi g_s$ is different from 
the one used in \ci{BG} by  factor of 1/2).
In particular, we find the following string one-loop contribution to the entropy, 
\be \delta  S=\  {5\pi^2\over 256} \ (2 g^2_{\rm YM} N)^{1/2}\ V_3\ T^3 \ .
\ee
Since it comes from the one-loop correction  with the smallest
number of derivatives, this   should be the dominant term at
strong coupling. 
One-loop contributions to the entropy
that scale as $T^3$ were recently discussed in
\cite{BR}.

Another interesting question is whether the $SL(2,Z)$ symmetry
imposes simple constraints on the function $f(g_{{\rm YM}}^2 N)$.
In \cite{BG} it was pointed out that, although the underlying
theory is certainly S-dual, the term-by-term expansion of
the function of the 't~Hooft coupling does not satisfy any simple
constraints. Hence, there is no paradox associated with the
appearance of the fractional power of $g_{{\rm YM}}^2 N$ at strong coupling.
Indeed, in choosing a fixed 't~Hooft coupling we automatically
choose a small $g_{{\rm YM}}$ which masks the constraints
imposed by the $SL(2,Z)$ in the 't~Hooft limit (cf. \ci{eguchi}).

%%%%%%%%%%%%%%%%%%%%%%%%%%%%%%%%%%%%%%%
\section{Free energy of other CFT's  }
\label{OtherCFTs}
%%%%%%%%%%%%%%%%%%%%%%%%%%%%%%%%%%%%%%%%%%

The low-energy limit of the world volume theory of coincident D3
branes is in many ways of special importance 
because it is related to
four-dimensional Yang-Mills theory. However, there are other
interesting brane configurations related to conformal field theories
in other dimensions, and in this section we discuss their
thermodynamic properties.

%%%%%%%%%%%%%%%%%%%%%%%%%%%%%%
\subsection{D5+D1
 and black hole entropy}
%%%%%%%%%%%%%%%%%%%%%%%%%%%%%%%%%%

One such interesting configuration involves $N_1$ D1-branes bound
to $N_5$ D5-branes \ci{SV,cm}. This defines a conformal field theory in
$1+1$ dimensions whose coupling constant is measured by $g_s$. 
When $g_s N_1$  and $g_s N_5$ are
much smaller than one, then this field theory can be studied
using standard D-brane methods and its entropy may be
counted. In the limit $g_s N_1, g_s N_5 \rightarrow \infty$ the
conformal field theory describes a black string in six dimensions,
whose near-horizon geometry is (as in the S-dual case \ci{CT,TTT})
 the
$AdS_3 \times S^3\times T^4$ background of type IIB supergravity.
In \cite{SV} the entropy of BPS states at weak coupling was compared
with the Bekenstein-Hawking entropy, which is applicable
at strong coupling, and  perfect agreement was found. 
An obvious reason for this agreement is supersymmetry. In \cite{cm,HSS}
this agreement was extended to near-extremal entropy.
Here the supersymmetry argument cannot be used, and the 
exact agreement of the coefficients is puzzling.

Motivated by our analysis of the $\alpha'$ corrections, we
propose an explanation.  From the point of view
of string theory on $AdS_3 \times S^3\times T^4$, the leading corrections
in inverse powers of $g_s N_1$  and $g_s N_5$ should come from
the $C^4$ terms, in complete analogy with our $AdS_5\times S^5$
calculation. The answer is particularly simple here
because the Weyl tensor
vanishes even for the near-extremal solution!  The space-time
contains a  product of two three-dimensional spaces, 
which separately cannot have
any Weyl curvature. It turns out that the Weyl tensor for the whole
space factorizes
(and therefore vanishes) only if 
the radii of the $AdS_3 $ and $S^3$ factors are {\it equal}.
This requirement is fulfilled here, just as in the
$AdS_5\times S^5$ case.

Furthermore, if we assume that all the $\alpha'$ 
corrections can be written in an appropriate scheme in terms of the
Weyl tensor and superpartners,\foot{For related discussion of 
higher-order terms in 
the string effective action and possible subtleties see
\ci{HTT}.}
  then it would seem that the action and
the geometry are not corrected, even away from
extremality.\foot{This is not so surprising
when the $5+1$  bound state is realized in terms of solitonic five-branes and
fundamental strings \ci{TTT}: the (orbifold of) $SL(2,R) \times SU(2)$ 
 WZW model is  an exact conformal theory. 
See also \ci{DAL} for a related discussion.} 
Indeed, the  argument is much strengthened by observing that
in the near-extremal limit  the  $r_0$-deformed $AdS_3$ factor is actually
\cite{HYU,jthroat,STR} the  BTZ black hole \ci{BTZ} 
 which  is locally  still 
 $AdS_3$. Thus even for  $r_0 \not=0$
we have locally $AdS_3 \times S^3$,  i.e. a space
{\it  conformal to a flat  space}
 because of equal radii of the factors.

 This implies  
that all $\a'$
corrections    vanish,  and thus   the free energy in the CFT of
the D1-branes bound to the D5-branes  is
completely independent of the string coupling!
Thus, we have found a plausible explanation for the
agreement of the near-extremal entropy of the 6d black string
with the weakly coupled D-brane calculation.
Let us also note that a completely analogous argument 
applies to near-extremal black holes in $D=4$ whose near-horizon
geometry is $AdS_2\times S^2$, and facilitates string theoretic
studies of their entropy. Black holes of this type include
the classic Reissner-Nordstrom solution \cite{KK,LS,CT}.\foot{The 
above argument  applies also to the rotating versions of the $D=5,4$
black holes  (for a discussion of their near-horizon geometry
and entropy see \ci{CVL}).}

Our line of reasoning relies completely on having the non-extremality
parameter $r_0$ much smaller than the scale of the $AdS_3$ geometry,
because only the near-horizon limit of the geometry decomposes into a
product of three-dimensional or flat factors.  One would therefore
expect any corrections to the entropy of the full D1+D5 solution
(which no longer enjoys this product structure) to contain inverse
factors of the extremal mass.  

%%%%%%%%%%%%%%%%%%%%%%%%%%%%%%%%%%%%%%%%
\subsection{Free energy of M-branes}
%%%%%%%%%%%%%%%%%%%%%%%%%%%%%%%%%%%%%%%%%%%%%%

We now discuss the 
conformal theories on multiple coincident M5 and M2 branes. 
Because of the conformal invariance, the higher-order corrections to the $D=11$ supergravity action  are expected to change 
 the leading-order expressions  \rf{eee}  for the corresponding free energies 
\ci{ENT} 
 only  by replacing the
 coefficient $k_p N^{p+1 \ov 2}$ by  a function $f_p(N)$  which approaches
$ k_p N^{p+1 \ov 2}$ for  large $N$. 
  In the case $D=10$,
$p=3$, discussed in the previous two sections, there were two
independent dimensionful quantities ($\alpha'$ and the Planck length)
which could be combined in appropriate power with the radius of
curvature of the geometry to obtain dimensionless parameters ($g^2_{\rm YM} N$
and $N$) which parametrize these subleading corrections.  Now the
situation is somewhat different because there is no $\alpha'$
parameter in M-theory, only the Planck length $l_{11} \sim
\kappa^{2/9}_{11}$.
  Correspondingly, we shall demonstrate explicitly
that there are only $1/N$ corrections to the free energy.
This will be done by  performing
the direct analog of the computation of the leading correction
\rf{coo} to the D3-brane free energy in the case of M5 and M2 branes.
We will not work out the perturbed geometry in these cases, 
though this is straightforward using 
the same methods  as in section~\ref{PerturbedSolution}.

The $D=11$  supergravity has a cubic  $R^4$ 
1-loop  UV divergence \ci{FTT} which, when computed  
with  a specific  cutoff $\Lambda_{11} = \pi^{2/3} l_{11}^{-1}$
 motivated by string theory \ci{GGG,RRT},
leads to the following correction to the leading Einstein  action 
\be I = - 
\int d^{11} x \sqrt g\  \bigg( {  1 \ov 2 \kappa^2_{11} }R +  
 { 1 \ov \kappa_{11}^{2/3}}\ \xi\   W + ...\bigg) \ , \ee
$$
\kappa_{11}^2 = 2^4 \pi^5 l^9_{11} \ ,
% 2^7 \pi^9
\ \ \ \ \ \ \    \xi ={2\pi^2 \ov  3}  \ , 
$$
where  
$W \sim  RRRR$  has the same structure as \rf{rrer}.
The coefficient of this  $R^4$ term is expected to be universal
(as it may be related to the  `anomaly' 
 $\int C_3 R^4$ term \ci{DUU} by supersymmetry \ci{GGG}).

The near-horizon limits of the extremal M2 and M5 solutions, i.e.
the backgrounds \ci{gt}
$(AdS_{7})_{2L}  \times (S^{4})_{L} $    and 
$(AdS_{4})_{\half L} \times (S^{7})_{L} $  (where we 
have indicated the values of the radii of the factors)
have maximal supersymmetry  and should not be 
modified by higher-order corrections to the effective action of 
 M-theory \ci{KR}. Since these $D=11$ spaces 
are not conformally flat (because of 
the different radii of the two factors),  their exactness should be manifest
in a  `$D=11$ supersymmetric'  scheme  which may be 
different from the one 
in which  $W$ is expressed in terms of the Weyl tensor  only \rf{rrrr}.
We shall assume that such scheme in which extremal 
solutions are exact  indeed exists and concentrate on corrections to 
the non-extremal ($r_0$-dependent, cf. \rf{throatmetric})
solutions.

The  important  observation  that  allows us to use 
the $(p+2)$-dimensional analogue of
the expression \rf{rrrr} to compute  corrections to the free energy
of Mp-branes is that  the sphere ($S^{9-p}$)
part of the  metric and  the 4-form field strength of 
the near-horizon backgrounds 
are {\it not}  modified by the non-extremality parameter $r_0$. 
As a result,  the $r_0$ - dependent correction to the action
is effectively determined by  the 
$p+2$ dimensional ($S^{9-p}$ compactified) theory.
Since $AdS_{p+2}$ is conformally flat (and, by assumption, 
 must not be modified
by the $R^4$ term), the 
correction is again  described 
by the Weyl-tensor dependent  term  \rf{rrrr}  only.
\foot{As in the D3-brane case, we  are ignoring  possible 
additional terms depending on the  $F_4$  field strength.
We  expect that  such terms which have the same dimension
as  $R^4$  and have  `reducible'  contractions of indices only
are scheme-dependent and thus are adjusted 
to have $AdS_{p+2} \times S^{9-p}$  as an exact solution, 
while `irreducible' terms vanish in the case of 
our backgrounds which have $r_0$-independent   $S^{9-p}$
and $F_4$  parts.}

%%%%%%%%%%%%%%%%%%%%%%%%%%%%%%%%%%%%%%%%%%
Let us start with   the M5-brane case.
The throat  limit of black M5-brane metric  (see, e.g., \ci{DLP})
is 
\be ds^2= {r\over L} (- f dt^2 + \sum_{i=1}^5 dx_i^2)+
{L^2\over r^2} f^{-1}dr^2  + L^2 d\Omega_4^2
\ ,
\ee
where $f= 1- {r_0^3\ov r^3}$.
Introducing the variable $y= \sqrt{L r}$, 
we bring the metric into the form
used in \ci{newWit} to describe a black hole in $AdS_{p+2}$
(cf. \rf{throatmetric})
 \be
 ds^2= {y^2\over L^2} \bigg[- (1 - {y^6_0 \ov y^6} ) dt^2 +  \sum_{i=1}^5 dx_i^2\bigg]+
4 {L^2\over y^2} (1 - {y^6_0 \ov y^6} )^{-1}
  dy^2  + L^2 d\Omega_4^2
\ , \ \ \ \ \ \   y_0 \equiv  (L r_0)^{1/2} \ .  \la{yy}
\ee
Computing the Weyl tensor for the 
7-dimensional  part of this metric
one finds that the 
 invariant  in \rf{rrrr} 
has the following value\foot{This computation  can be done   using
either
\rf{GenMet},\rf{GenW} or, directly, the  curvature  of \rf{yy}.
The Euclidean   7-metric has  the 
 Ricci tensor
$
R_{mn} = -{ 3\ov  2 L^2} g_{mn}  , \ \ R= -{ 21\ov 2 L^{2}}   , 
$
and the  Weyl tensor $(i=1,..,5)$ \\
$
C^{\tau y}_{\ \ \tau y} =   { 5 \ov 2}  X  , \ \ 
C^{\tau i}_{\ \ \tau j} =  - { 1 \ov 2}  X  \delta^i_j    ,  \ 
C^{yi}_{\ \ y j} = -  { 1 \ov 2}  X   , \ \ 
C^{ij}_{\ \ kl} =    { 1 \ov 4} X ( \d^i_k \d^j_l - \d^j_k \d^i_l)  , 
\ \ 
X\equiv { 1\ov L^2} {  y_0^6\ov  y^6}.$}
\be
W=\  { 3285 \ov 64 L^8} \ { y_0^{24} \ov y^{24}} \ . 
\ee
The corresponding correction to the  Euclidean action  then takes the form
\be
 \delta I =\   - { 3285 \ov 64 L^8}\ 
 \xi \  \kappa_{11}^{-2/3}\  {\rm Vol} (S^4) \ \beta\  V_5  \ 
\int_{y_0}^\infty dy \sqrt { g_7} \ 
  {y_0^{24 }\ov  y^{24}}\ ,
\ee
where ${\rm Vol} (S^4) = {8\pi^2 \ov 3} L^4$ and 
  $\sqrt {g_7}  = 2 {y^5\ov L^5}$. 
 As follows from the condition of regularity of the Euclidean metric,
the temperature is 
\be T= \beta^{-1} = {3y_0\ov  4\pi L^2}
\ .
\ee
As a result, the 
leading correction to the free energy is given by
\be
 \delta F  =\  - \ k  \kappa_{11}^{-2/3} L^3 V_5 T^6 \ , \ \ \ \ \ \ \
k =  {8 \pi^2 \ov  27}\cdot \xi \cdot ({4\pi\ov 3})^6 \cdot { 3285 \ov 64}    \ .
\ee
Using  the charge quantization for M5 branes we find
the relation \cite{kleb}
\be 
L^9= N^3 {\kappa_{11}^2\over 2^7 \pi^5}\ .
\ee
Substituting it into $\delta F$, we finally obtain
\be  
 \delta F  =  -  a_1   N V_5 T^6 \ , 
\ \ \ \ \ \  \ee
$$   a_1  = \ 730\ ({2\pi\ov 3})^8\  ({\pi\ov 2})^{1/3} 
 \ . 
$$
To this order the free energy of the  $d=6$ 
world-volume 
$(0,2)$ 
theory is thus  predicted to be 
given by
\be
 F = -  V_5 T^6\ (a_0 N^3 + a_1 N)
\ , 
\ee
where $a_0=2^6 3^{-7} \pi^3$ \ci{ENT} and $a_1 $
 are positive numerical coefficients.

In  the M2 brane case  the  throat  metric is
\be
 ds^2= {r^4\over L^4} (- f dt^2 + \sum_{i=1}^2 dx_i^2)+
{L^2\over r^2} f^{-1}dr^2  + L^2 d\Omega_7^2
\ , \label{MTwoThroat}
\ee
where $f= 1- {r_0^6\ov r^6}$.
The variable $y= {r^2\ov L}$ brings \eno{MTwoThroat} into the standard form
for a black hole in $AdS_4$:
\be
 ds^2= {y^2\over L^2} \bigg[ - (1- {y_0^3\ov y^3} ) 
 dt^2 + \sum_{i=1}^2 dx_i^2\bigg]+
{L^2\over 4 y^2} (1- {y_0^3\ov y^3} )^{-1}dy^2 
 + L^2 d\Omega_7^2
\  , \ \ \ \ \ \ \ \   y_0 \equiv {r_0^2 \ov L} \ . 
\ee
The product \rf{rrrr} 
of  the  Weyl tensors of the 4-dimensional part of this
 metric  is  found to be equal to (cf.\rf{GenW})\foot{This  Euclidean
4-metric has 
 the following Ricci tensor
$
R_{mn} = -{ 12\ov   L^2} g_{mn}  , \ \  R= -{ 48\ov  L^{2}} , 
$
and   Weyl tensor $(i=1,2)$\ 
$
C^{\tau y}_{\ \ \tau y} =   4  X  , \ \ 
C^{\tau i}_{\ \ \tau j} = - 2  X  \delta^i_j    ,  \ $ $
C^{yi}_{\ \ y j} =  -2  X    ,  \ \ 
C^{ij}_{\ \ kl} =    4 X ( \d^i_k \d^j_l - \d^j_k \d^i_l)  , \ \ 
X\equiv { 1\ov L^2} {  y_0^3\ov  y^3}  .
$} 
\be
W=\ { 1152 \ov L^8}\ { y_0^{12} \ov y^{12}} \  . 
\ee
The correction to the Euclidean  action is then 
\be \delta I=  \   - { 1152 \ov L^8}\  \xi \  \kappa_{11}^{-2/3}
 \  {\rm Vol}(S^7)\  \beta \  V_2 \ 
\int_{y_0}^\infty dy \sqrt {g_4}\  {y_0^{12} \ov L^8 y^{12} }
\ , 
\ee
where    Vol$(S^7) = {\pi^4 \ov 3} L^7$, 
\  $\sqrt {g_4}  = \half {y^2\ov L^2}$,
and  
 \be  T=\beta^{-1} = {3y_0\ov  2\pi L^2} \ . \ee
 The 
leading correction to the free energy is  thus
\be
 \delta F  = \   -\   m   
 \kappa_{11}^{-2/3} L^3 V_2 T^3 \ , \ \ \ \ \ \ \
  m= 1152\cdot  {\pi^4 \ov 18}   \cdot ({2\pi\ov 3})^5  
\ . 
\ee
Using the relation that follows from 
the M2 brane charge quantization \cite{kleb},
\be 
L^9= N^{3/2} {\kappa_{11}^2 \sqrt 2\over \pi^5}\ ,
\ee
we finally have
\be
 \delta F = - b_1  N^{1/2} V_2 T^3 \ , \ \ee
$$
b_1 =\  64 \ ({2\pi \ov 3})^5 \   2^{1/6}\  \pi^{7/3}
  \ . 
$$
To this order the free energy of the conformal theory on
$N$ coincident M2-branes is thus given by
\be F = - V_2 T^3\  (b_0 N^{3/2} + b_1 N^{1/2})
\ , 
\ee
where $b_0= 2^{7/2} 3^{-4} \pi^2$ \ci{ENT} and $b_1$ are positive.

%%%%%%%%%%%%%%%%%%%%%%%%%%%%%%%%%%%%%%%%%%%%%%%
We end this section with several comments.
It  was suggested  in \ci{RRT} that not only $R^4$ but all 
 the terms
  $R^{3n+1}$ ($n=1,2,...$) 
 may play a similar  special role in  M-theory.
In particular, on dimensional grounds, 
they appear in the effective  action multiplied by 
{\it integer} powers of the inverse M2-brane  tension, 
$T_2^{-1} = 2\pi l^{3}_{11}$.
Including such terms and 
repeating the above analysis
we find the following generalizations
of the expressions for the free energy in the M5-brane case 
\be F = -  V_5 T^6 N^3\bigg(a_0   + \sum^\infty_{n=1} {a_n \ov  N^{2n}} \bigg)\  , 
\ee
and  the M2-brane case 
\be F = -V_2 T^3  N^{3/2} \bigg(b_0  + \sum^\infty_{n=1} {b_n \ov  N^{n}} \bigg)\  , 
\ee
where $a_n$ and $b_n$ are numerical coefficients.  Although we expect
$a_1$ and $b_1$ to be given reliably by the first-order perturbation
to the action, the perturbed geometry will almost certainly
affect the $a_n$ and $b_n$ with $n>1$.

In the case of $SU(N)$ SYM theory, comparison with standard field
theoretic methods may become possible if one manages to
extrapolate the series in inverse powers of the `t~Hooft coupling,
calculated from type IIB string theory, to weak coupling.
Is there an analogue of this statement for the CFT's on M-branes?
Here, the only parameter is $N$ and the coupling cannot be dialed
separately.  The details of the field theory are rather poorly known for
$N> 1$.  
The M2-brane CFT is expected to be effectively described 
by an IR  fixed point of $D=3$, $U(N)$  SYM, while the M5 brane theory --
by an UV fixed point of $D=5$, $U(N)$  SYM.  
The world volume theory becomes free in the conformal limit for $N=1$
and the coefficient in the free energy can be fixed at this point.
However, even if we knew the full asymptotic $1/N$ expansion of the
free energy in supergravity, it is not clear that a reliable
comparison could be made at $N=1$.

\section*{Note Added}

After this paper was completed a question was raised \cite{pt}
concerning the precise meaning of the metric derived in 
section 3.\foot{We are grateful to S. Theisen for sending 
us  the draft of their paper prior to publication.}
Here we clarify this issue. We consider the following ansatz for the
10-dimensional metric in the Einstein frame,
\be  ds^2_{10}  = e^{-{ 10 \over 3}\nu (x)} g_{5mn} (x) dx^m dx^n
 + e^{2 \nu(x)} d\Omega_5^2
\ ,\ee
where we set $L=1$. Taking the standard ansatz for the 5-form field
and compactifying on $S^5$, we find the following 5-dimensional
effective action,
\be
I_5 =-{1\over 16\pi G_5}
\int d^5 x \sqrt{g_5} \bigg[ R_5 - { 1 \over 2} (\partial \phi)^2
- { 40 \over 3}(\partial \nu)^2 -  V(\nu)
 + \gamma    e^{10 \nu - {3\ov 2} \phi} \bigg(W   + O( (\del \nu)^2) \bigg)
\bigg]
\ ,
\ee
where
$$
V(\nu) = 8 e^{-{ 40 \over 3}\nu} -20 e^{-{ 16 \over 3}\nu}  \ . 
$$
This shows that the field $\nu$, which is the logarithm of the
5-sphere radius, acquires a mass; hence, it is a ``fixed scalar.''
If $\g=0$,  then the classical value of $\nu$ can be chosen to be zero.
This field receives a source of order $\g$ and the classical
equation becomes
\be \nabla^2 \nu   -  32 \nu   + { 3 \over 8}\g  W  +
O(\nu^2) +   O(\g^2) =0
\ .\ee
Its solution is \cite{pt}
\be \label{vari}
\nu = {15\g\over 32} {r_0^8\over r^8} \left (1+ {r_0^4\over r^4} 
\right
) + O(\g^2)
\ .\ee
The 5-dimensional metric $g_{5mn}$ is precisely what we determined
in section 3  (both the metric and the dilaton corrections  
do not  mix with $\nu$ to leading order in $\g$).  
This is the perturbed metric of a black hole in $AdS_5$
which governes the dynamics of fields which do not depend on
the 5-sphere coordinates. However,
the variation of the 5-sphere radius in the
10-dimensional metric, given by (\ref{vari}), does affect
the modes that come from higher Kaluza-Klein harmonics.

%%%%%%%%%%%%%%%%%%%%%%%%%%%%%%%%%%%%%%%%%%%%%%%
\section*{Acknowledgements}
%%%%%%%%%%%%%%%%%%%%%%%%%%%%%%%%%%%%%%%%%%%%%%%%%%%%
We are grateful to E. Kiritsis, 
H. Liu, A. Rajaraman and E. Witten for discussions and 
to Yu.N. Obukhov for his kind help with the 
GRG computer algebra system \ci{ZZ}.
The work  of I.R.K. and S.S.G. 
was supported in part by US Department of Energy 
grant DE-FG02-91ER40671 and 
by the James S. McDonnell
Foundation Grant No. 91-48.  S.S.G. also thanks the Hertz
Foundation for its support. 
The  work  of A.A.T. was supported in part
by PPARC,   the European
Commission TMR programme grant ERBFMRX-CT96-0045
and  the INTAS grant No.96-538.

%%%%%%%%%%%%%%%%%%%%%%%%%%%%%%%%%%%%%%%%%%%%%%%%%%%%%%%%%%%%

\end{document}